\documentstyle[12pt]{article}
\begin{document}
\noindent
{\LARGE{\bf Investigation of conduction band structure, electron
       scattering mechanisms and phase\\
       transitions in indium selenide 
       by means of transport measurements under pressure}}
				    
\vspace{0.3in}
\begin{center}
{\large
D. Errandonea, A. Segura, J.F. S\'anchez-Royo, V. Mu\~noz 

Dpto. de F\'{\i}sica Aplicada (Facultad de F\'{\i}sica),
Universidad de Valencia,
C/Dr. Moliner 50, E-46100 Burjasot (Valencia), Spain
	      
\vspace{0.15in}                       
P. Grima, A. Chevy

Laboratoire de Physique des Milieux Condens\'es, 
Universit\'e Paris VI\\
F-75251 Paris Cedex 05, France
				     
\vspace{0.15in}
C. Ulrich
	      
Max-Planck Institut f\"ur Festk\"orperforschung, Sttutgart\\
D-70569 Stuttgart, Germany}
\end{center}

\vspace{0.3in}
\begin{abstract}
In this work we report on Hall effect, resistivity and thermopower 
measurements in n-type indium selenide at room temperature 
under either hydrostatic and quasi-hydrostatic 
pressure. Up to 40 kbar (= 4 GPa), the decrease of  carrier 
concentration  as the pressure increases is explained
through the existence of a subsidiary minimum in the conduction band. 
This minimum shifts towards lower energies under pressure, 
with  a pressure coefficient  of about -105 meV/GPa, and its related
impurity level traps electrons as it reaches the band gap and 
approaches the Fermi level. The pressure value at which the 
electron trapping starts is shown to depend on the electron 
concentration at ambient pressure and the dimensionality of the 
electron gas. At low pressures the electron mobility increases under 
pressure for both 3D and 2D electrons, the increase rate being 
higher for 2D electrons, which is shown to be coherent with 
previous scattering mechanisms models. The phase transition from 
the semiconductor layered phase to the metallic sodium cloride phase 
is observed as a drop in resistivity around 105 kbar, but above 40 
kbar a sharp nonreversible increase of the carrier concentration is 
observed, which is attributed to the formation of donor
defects as precursors of the phase transition.
\end{abstract}
				    
\vspace{0.15in}

\pagebreak                                     
\section{Introduction}
\indent

Optical properties, lattice dynamics and structural phase 
trasitions of layered III-VI semiconductors under compression 
have been widely investigated. Pressure experiments have shown to 
be an idoneous tool to investigate the specificity of chemical 
bonds and the electronic structure in  these semiconductors. 
A large variety of interesting properties and behaviours have been 
so detected: large anisotropy and  nonlinearity in the elastic 
properties\cite{1,2,3}, nonlinearities in the  pressure dependence
of phonon frequencies\cite{2,4,5,clemenz} and  electronic 
transitions\cite{2,6,7,8,9},  layered phase
instabilities\cite{3,4,10}, etc.

Unlike optical and lattice dynamical properties, transport 
properties of III-VI semiconductors under pressure have been 
little investigated.  Ismailov {\it et al}\cite{11} have 
reported on Hall effect (HE) and resistivity measurements 
in n-type indium selenide (InSe) up to 10 kbar. These 
authors directly relate the pressure dependence of the carrier 
concentration to the pressure coefficient of the direct band gap. 
Such a direct correlation can hardly be expected in samples 
exhibiting a clearly extrinsic behaviour. 

In the present paper, we report on a systematic study of transport 
properties under high pressure, at room temperature (RT),
in the  layered  III-VI semiconductor indium selenide 
doped with different donor impurities. Experimental methods are 
described in Section II. Results on HE, resistivity and 
thermopower experiments under pressure are detailed in Section III.  
Section IV is devoted to the discussion of results and its 
implications concerning  the conduction
band structure, electron scattering mechanisms and layered phase 
stability in InSe. The pressure dependence of the transport
parameters below 40 kbar is consistent with a previous model in
which both two-dimensional (2D) and three-dimensional (3D)
electrons are considered\cite{18}.

\section{Experimental}
\indent

The InSe crystals were prepared by the Bridgman method from a 
non-stoichiometric melt
In$_{1.05}$Se$_{0.95}$\cite{12}. The tin (Sn) or silicon (Si) 
impurities were introduced in the preparation of the 
polycrystalline melt as selenium compounds\cite{12,13}. It has 
been shown\cite{13} that only a small part of impurities remains in 
the crystal, the rest being rejected to
the end of the ingot with the indium excess. Neutron transmutation 
doped (NTD) InSe crystals have also been analyzed in this work. The 
preparation, transport
characterization and defect annealing of these crystals, whose Sn 
contents depend on
the neutron dose, have been reported elsewhere\cite{14,15}. Samples 
were cleaved from the ingots with a razor blade and cut into
parallelepipeds; the thickness and 
shape  of the samples were different for each kind of experiment. 
All the measurements were performed at RT. 

\subsection{Resistivity and Hall effect measurements in 
hydrostatic pre\-ssure up to 12 kbar}
\indent

Samples for these measurements were 10 to 30 $\mu$m thick and about 
3$\times$3 mm$^2$ in size. Contacts were made in the Van der Pauw 
configuration\cite{16} by soldering with high-purity indium.  
Resistivity measurements under pressure were carried out in a 
conventional maraging steel vassel, with pentane as   pressure 
transmitting fluid.  Moreover, HE and resistivity measurements 
were performed in a Unipress copper-berillium  cell with the 4:1 
methanol-ethanol mixture as pressure transmitting fluid. 
Pressure was measured in both cases with a manganine gauge.

\subsection{Resistivity measurements in quasi - hydrostatic 
pressure up to 110 kbar}
\indent

Samples for these measurements were parallelepipeds 50 to 70 $\mu$m 
thick and about 2$\times$2 mm$^2$ in size. Indium contacts were made
in the Van der Pauw configuration by vacuum evaporation. The 
pressure cell was a large volume V2 cell\cite{17} with a pair
of tungsten carbide Bridgman anvils of 10 mm in diameter,  
with steel binding rings. The Bridgman 
gasket assembly and geometry are shown in the inset of Figure 3. 
Gaskets were made of pyrophyllite, the pressure transmitting 
medium was sodium cloride or silver cloride and the electric leads 
between both gasket were made of  copper foil
20 to 30 $\mu$m thick.  Pressure was previously calibrated with the 
phase transition points of a  bismuth gauge.

\subsection{Resistivity and Hall effect measurements under  
quasi - hydrostatic pressure up to 55 kbar}
\indent

Samples for these measurements were parallelepipeds 50 to 70 $\mu$m 
thick and about 5$\times$5 mm$^2$ in size. Ohmic contacts were made
in the Van der Pauw configuration by soldering silver wires to the 
high-purity indium contacts pads previously vacuum evaporated. In 
order to obtain good contacs the end of the silver leads was 
previously flattened. In this case, the pressure cell consist in a 
pair of Bridgman tungsten carbide anvils, 27 mm in diameter,  
without steel binding rings.  Besides the gasket thickness and 
diameter, and the pressure medium, which
was either sodium cloride or boron nitride charged epoxy, the 
geometry was  similar to that of the previous method.  

The initial thickness of the pyrophyllite gaskets was 0.5 mm and
the diameter of the hole was 9 mm. They were treated at
680$^o$C during one hour for getting suitable mechanical 
properties\cite{piro} and avoiding the break of the gaskets in
the first stages of the compression. Before the prefiring, 
four channels were made in one of the gasket to place the silver 
wires. Bridgman anvils were put between the pistons of a 300 tonnes
press. The pressure was determined by calibration of the oil
pressure against the bismuth phase transition points.
A copper coil was placed around one of the pistons.   
 The magnetic field intensity in the gap 
between the anvils was measured by means of a Hall effect probe, 
as a function of the distance between them, and was 0.6 T at
the normal working distance (0.8 mm).

\subsection{Thermopower}
\indent

Thermopower measurements under pressure were carried out in the 
system described in subsection 2.2. The geometry of 
the gasket and sample assembly is shown in the inset of Figure 6.  
One point of the sample is heated by Joule effect in a thin 
constantan strip in thermal contact with it and soldered to Cu 
leadtrhoughs. In high pressure conditions a stationary thermal 
gradient is quickly attained due to the thermal conduction trough 
the anvil-gasket assembly .
Two strip  Cu/Constantan thermocouples are used to measure both 
the temperature gradient and the Seebeck voltage 
in the sample.
 
\section{Results}

\subsection{Resistivity measurements in hydrostatic pressure}
\indent

Figure 1 shows the resistivity ($\rho$) as a function of pressure for 
several InSe samples. Five of them (curves 1 to 5) were chosen as 
representative for well characterized three-dimensional behaviour 
in their transport properties\cite{14,18,19}. For concentrations
lower than 10$^{16}$ cm$^{-3}$ a slight decrease of the
resistivity is observed between ambient pressure and 8 kbar
followed by a slight increase up to 12 kbar (sample 1). The rest
of those samples exhibit 
a regular trend: the resistivity increase under pressure 
is more pronounced for higher electron concentrations ($n$). 
On the other side,  samples with well characterized two-dimensional 
behaviour in their transport
properties\cite{19} exhibit a resistivity decrease under pressure 
(curves 6 and 7). 

\subsection{Hall effect under hydrostatic pressure}
\indent

Figures 2-a and 2-b show the electron concentration and 
mobility ($\mu$), respectively as a function of pressure up to 
12 kbar for Si-doped, Sn-doped  and NTD InSe samples. Within this 
range of pressure the electron mobility monotonically increases 
with pressure in all of them. The resistivity increase under 
compression in the Sn-doped and NTD samples is then mainly due to 
the decrease of the carrier concentration. In the other side, in 
Si-doped samples the electron concentration slighty increases with 
pressure. Thus, the increase of both the mobility and the carrier 
concentration produces the observed decrease of the resistivity 
(Figure 1). It must be outlined that the apparent electron 
concentration at zero pressure is very similar in one of the 
Sn-doped samples ($\circ$) and one of the Si-doped samples 
($\triangle$). Nevertheless, their transport properties 
($\rho$, $n$ and $\mu$) have a completely different behaviour under 
compression. Then, pressure effects in the transport properties of 
InSe turn out to be very sensible to the dimensionality of charge 
transport, which was established through independent experiments 
(resistivity anisotropy, photo Hall effect, etc,)\cite{18}.

\subsection{Resistivity measurements under quasi-hydrostatic
pre\-ssu\-re up to 110 kbar}
\indent

Figure 3 shows the resistivity as a function of pressure for 
a Sn-doped and a Si-doped InSe sample, with an effective electron 
concentration of 10$^{17}$ cm$^{-3}$ and 6 $\times$ 10$^{16}$ 
cm$^{-3}$ at ambient pressure, respectively. It can be seen there 
that, above 10 kbar in the Sn-doped sample and above 20 kbar in the 
Si doped one, the resistivity exponentially increases up to around 
40 kbar. It reaches a maximum centered at 
about 45 kbar and then monotonically decreases  up to 100 kbar. An 
abrupt decrease is also observed at
around 105 kbar. It must be outlined that changes over  the 
resistivity maximun are not reversible. Points of curve 2 were 
taken for decreasing pressures.

\subsection{Hall effect and resistivity measurements under 
quasi - hy\-dros\-tatic pressure}
\indent
     
Figures 4 and 5 show the resistivity, the electron concentration, 
and Hall mobility as a function
of pressure for different InSe samples doped with 
Sn and Si, respectively. In the range of pressure up to around 12 
kbar, these results confirm those of Figures 1 and 2. Having the
electron concentration a weak pressure dependence up to 25 kbar
in Si-doped samples, the resistivity decrease turns out to be a 
direct consequence of the electron mobility 
increase under compression. Above that pressure, both the carrier
concentration and the mobility 
have a slow decreasing behaviour providing an increase
of the resistivity. On the contrary, above 12 kbar in the Sn-doped 
samples the electron concentration decreases by one order of 
magnitude and the mobility by factor 3 to 5. 
In addition, from Figure 4 it is clear again that in samples
with 3D behaviour the increase rate of the resistivity is stronger 
for those with higher electron concentrations at zero pressure. 
Notice that, in one of the Sn-doped samples ($\circ$) 
the pressure range could be extended up to 55 kbar,
i.e. above the resistivity maximum of Figure 3. The resistivity 
decrease above 45 kbar appears to be associated to a sharp 
increase of the electron concentration.

\subsection{Thermopower measurements under quasi-hydrostatic 
pre\-ssure}
\indent

Figure 6 shows the thermopower as a function of pressure for a 
Sn-doped InSe sample. The thermopower decreases with lowering
pressure in all the measured range and this 
decrease is sharper in the pressure range around 40-50 kbar. At 
the highest pressure, the thermopower value (about 10 mV/K) is 
typical of a degenerate semiconductor, coherently with the low 
value of resistivity observed in this pressure range (Figure 3).

\section{Discussion}

\subsection{Conduction band structure}
\indent

Hall effect results in InSe samples with well characterized 3D 
behaviour show that the electron concentration decreases as 
pressure increases up to some 40 kbar. This behaviour cannot 
be attributed to the pressure change of the band gap
(the electron concentration remains in the extrinsic range) 
and neither to the change of the ionization energy of the tin 
related shallow donor, which is fully ionizated at RT. 
At ambient pressure this ionization energy is 17.4 meV\cite{20}, 
which is very close to the ideal hydrogenic donor in
InSe. The hole effective mass being very high in InSe, 
the pressure change of the
shallow donor ionization energy should be very similar to 
that of the exciton binding energy, that
has been shown to decrease under compression\cite{9}. 
That effect, also observed in GaSe\cite{2},  was atributed 
to the increase of the static dielectric constant under 
pressure. This hypothesis has been verified in the related 
compound GaS\cite{21} and recently in GaSe\cite{yo}.  

The decrease of the electron concentration under pressure has
been observed in III-V semiconductors and explained through the
existence of subsidiary minima in the conduction band moving to
lower energies with respect to the absolute minimum ($L$ and $X$
minima with respect to the $\Gamma$ minimum in GaAs). According
to band structure calculations\cite{band1,band2} there exist
subsidiary minima in the conduction band of InSe located a few
hundreds meV above the absolute one.
In the related compounds  GaS and GaSe these minima are
responsible for the indirect absorption edge, whose pressure 
coefficient are of the order of some -100 meV/GPa\cite{2,7}.  
In the case of InSe, the pressure behaviour of the 
exciton peak,  which
broadens and desappears above 25 kbar, has been explained 
through the interaction
with an excited minimum of the conduction band, 
moving to lower energies under pressure\cite{9}.
The decrease of the electron concentration under compression is
coherent with the assumption that, related to the excited
minimum, there exist a localized level resonant with the
conduction band at ambient pressure.
 As this excited minimum moves to lower energies with the pressure
increase, the related impurity level reaches the forbidden band 
before the semiconductor becomes indirect. Then it traps 
electrons as it approaches the Fermi level ($E_F$). If we assume
that all hydrogenic levels are ionized and the 3D electron 
concentration ($n_3$) decrease is due to trapping, 
the free electron concentration and the trapped electron 
concentration ($n_T$)
as a function of pressure ($P$) are given by:

\begin{eqnarray}
n_3(P)=N_C(P) e^{-\frac{E_c(P) -E_F(P)}{K_BT}}
\end{eqnarray}

and

\begin{eqnarray}
n_T(P)=n_3(0)-n_3(P)=N_D-n_3(P)=\frac{N_T}{1+\frac{1}{2}
e^{\frac{E_T(P) -E_F(P)}{K_BT}}}\quad,
\end{eqnarray}

\noindent
where $N_D$ is the shallow donor concentration, $N_T$ is the
deep impurity level concentration, $T$ is the absolute temperature, 
$K_B$ is the Boltzman constant, and $E_C$, $E_F$ and $E_T$ are the 
conduction band absolute minimum, the Fermi level, and the excited 
minimum related impurity level energies, respectively.   
If we assume that all impurity levels are associated 
to the tin impurities, we can make $N_T=N_D$ and, neglecting the 
pressure change in the effective density of
states $N_C$, we can obtain, from Equations (1) and (2), the
pressure dependence of $E_F$ and $E_T$ 
with regards to the absolute minimum of the conduction band. 
Figure 7 shows the results of this model.
A least square fit to $E_T$ - $E_C$ data with a cuadratic
function (dashed line in Fig. 7) gives:

\begin{eqnarray}
E_T - E_C = 100(15) - 10.5(2)P (\frac{\mbox{meV}}{\mbox{kbar}})+ 
0.04(0.02)P^2 (\frac{\mbox{meV}}{\mbox{kbar}^2}) \quad, 
\end{eqnarray}

\noindent
where $P$ is in kbar. Thus, at ambient pressure and temperature
the deep level lies at some 100 meV above the absolute minimum
of the conduction band. Furthermore,  
the pressure coefficient of the excited minimun impurity 
level turns out to be -105 meV/GPa, i.e. very close to the 
pressure coefficients of indirect absorption edge in GaS and GaSe.

\subsection{2D electrons}
\indent

The presence of 2D electrons in InSe and the role that they play 
in transport properties has been widely investigated
\cite{kress1}-\cite{howell}. The origin of 2D electrons 
in InSe is extrinsic. The 2D states are electric subbands 
that have been atributed to quantum size effects in thin layers 
limited by stacking fault related barriers\cite{18}. Other
authors have  proposed a model in which 2D electronic subbands in 
InSe are created by a high concentration of donor impurities bound 
to a stacking fault,  like in $\delta$-doping systems\cite{howell}. 
This model is inconsistent with the low  areal concentration and
high 2D electron mobility in InSe at low temperature. Below 20 K, 
transport properties are dominated by 2D electrons in most n-type 
InSe samples.  At RT, among samples with effective 
electron concentrations of the order of 10$^{17}$ cm$^{-3}$ , only 
Si-doped InSe exhibits  2D behaviour, which can been attributed to
the lower solubility of Si in InSe with respect to Sn, which makes 
a part of Si atoms to deposit in interlayer positions bound to the 
stacking fault, so increasing the 2D electron concentration.

It is to be outlined that transport  measurements under pressure 
give a suplementary evidence of the specificity of  Si-doped 
samples. This behaviour can be interpreted in the framework of 
the subband model proposed in Reference \cite{18}.
Between planar defects, to which 2D electrons are associated, 
the 3D electron concentration in Si 
doped InSe remains relatively low: from far infrared 
absorption\cite{20}, the  concentration  of Si-related
shallow donor was estimated  to be of the order of 5 - 6 $\times$
10$^{15}$ cm$^{-3}$. 
It means that the Fermi level at RT is some 120 meV below the 
conduction band.  Then, these samples behave as 3D samples with a 
low electron concentration. The deep
center related to the excited minimum in the conduction band does 
not trap 3D electrons because the Fermi level is deep in the 
forbidden band. On the other hand, electron in the electric 2D 
subbands are degenerate; owing to this characteristic, most of them 
occuppy energy states below the Fermi level. In this way, 2D 
electrons are not trapped until the deep level 
crosses the Fermi level, which occurs at much higher pressures. 
If we assume that the change of resistivity is mainly due to the 
decrease of the 2D electron concentration ($n_2$)
due to trapping, we can also try to estimate the 
pressure coefficient of the trap with respect to the bottom 
of the 2D electric subband ($E_{C2}$).  
The  electron concentration is related to the Fermi 
level through the well known equation\cite{stern}:

\begin{eqnarray}
n_2(P)=\frac{m^*K_BT}{\pi \hbar^2} 
\ln\left(1+e^{\frac{E_F(P)-E_{C2}(P)}{K_BT}}\right)\quad.
\end{eqnarray}

On the other side, 2D electrons in InSe are extrinsic. They are
provided by donor centers located between stacking faults. Their
energy level related to the absolute minimum is shallow and is
affected by quantum confinement effects, that are the origin of
the 2D subbands according to the model proposed in Ref.
\cite{18}. On the opposite, their level related to the second
minimum would be deep and localized and then it is not affected
by the quantum confinement effects, behaving then as electron
trap as it crosses the Fermi level under pressure.
We can consider, like in the 
previous subsection, that the trap concentration is the 
total number of 2D electrons at ambient pressure. Then, the  
concentration of trapped electrons in
each 2D defect ($n_{T2D}$) will be given by:

\begin{eqnarray}
n_{T2D}=n_2(0)-n_2(P)=
\frac{n_2(0)}{1+\frac{1}{2}e^{\frac{E_T(P)-E_F(P)}{K_BT}}}
\end{eqnarray}
	 
From Equations (4) and (5) we can obtain the pressure change of the 
trap level with respect to the bottom of the subband, that is shown 
in Figure 7. The pressure coefficient turns out to be -110 meV/GPa, 
of the same order that the one obtained from results in samples 
with 3D behaviour.
     
\subsection{Electron scattering mechanisms}
\indent

The pressure dependence of the electron Hall mobility provides 
us with a suplementary test of electron scattering mechanisms 
in InSe. As regards 3D electrons in the absolute minimum of the 
conduction band, the electron mobility ($\mu_3$) in InSe
at room temperature has been shown to be controlled by both polar 
and homopolar phonon scattering and by ionized impurity
scattering mechanisms\cite{18}. For scattering by polar phonons
an iterative method must be used\cite{nag}. The Schmid-Fivaz 
relaxation time\cite{fivaz} for homopolar phonon scattering 
and the Brooks-Herring relaxation time\cite{brooks} for 
ionized impurity scattering 
can be introduced in the elastic term of
the scattering rates in the iterative method.

The scattering rates of 3D electrons depend on the phonon 
energies, through the phonon occupation number ($N$), and on the 
corresponding electron phonon coupling constant. In the case 
of homopolar optical phonon scattering, the electron phonon 
coupling constant is:

\begin{eqnarray}
g^2=\frac{{D_0}^2 {m^*}^{\frac{3}{2}}}{2\sqrt{2}\pi N \hbar
(\hbar \omega_{hp})^{\frac{3}{2}}}
\end{eqnarray}

\noindent
where $D_0$ is the deformation potencial and
$m^* = ({m^*_{\perp}}^2m^*_{\parallel})^{\frac{1}{3}}$,
being $m^*_{\perp}$ and $m^*_{\parallel}$ the electron effective
mass perpendicular and parallel to the c-axis, respectively.
The value at atmospheric pressure of this coupling constant is 
$g^2$=0.028\cite{18} and the phonon involved is the $A'_1$
homopolar optical mode with wavelenght: $\omega_{hp}=$ 115
cm$^{-1}$\cite{gasanly}. 

In the case of polar phonon scattering mechanism, the coupling 
constant is the Fr\"ohlich constant, that is given by:

\begin{eqnarray}
\alpha=\frac{e^2 \sqrt{m^*}}
{\sqrt{2}\hbar \sqrt{\hbar \omega_{LO}}} 
\left(\frac{1}{\epsilon_{\infty}}-\frac{1}{\epsilon_{0}}\right)\quad,
\end{eqnarray}

\noindent
where $e$ is the electron charge and $\epsilon_{\infty}$ and
$\epsilon_0$ stand for the high-frequency and static
dielectric constants. 
As the electron scattering rates for LO phonons are obtained
through an integration over all the possible directions of the
phonon momentum, this constant has to be averaged over the whole 
solid angle. Assuming that its main angular dependence arises from
the variation of the dielectric constant\cite{lang}, its value
at room pressure is $\tilde{\alpha}=0.144$\cite{18}. The
wavelenght of the LO phonon is: $\omega_{LO}=$ 211
cm$^{-1}$\cite{gasanly}. 

The pressure dependence of the quantities involved in Equations
(6) and (7) are reasonably well known. The effective masses at
atmospheric pressure can be taken from Reference \cite{portal},
$m^*_{\perp}=0.141 m_o$ and $m^*_{\parallel}=0.081 m_o$ 
being $m_o$ the free electron mass. 
In the framework of the {\bf k.p} theory\cite{kane} their variation 
with pressure is proportional to the direct band gap ($E_g$) 
variation. The perpendicular component of static dielectric constant
$\epsilon_{0\perp}$ is nearly constant\cite{5} and the 
variation of $\epsilon_{0\parallel}$ can be obtained from the 
behaviour under pressure of the exciton binding energy ($R$)\cite{9}.
Thus, the decrease of $R$ imposes that $\varepsilon_{0\parallel}$
increases strongly from 7.8 to 10.9 up to about 40 kbar. This
behaviour has been also observed in related compounds\cite{2,21,yo} 
and it was explained through the charge transfer from intralayer to
interlayer space\cite{2}. Finally, the pressure dependence 
of $\epsilon_{\infty \parallel}$ and $\epsilon_{\infty \perp}$ can 
be calculated from that of $\varepsilon_0$ and the phonon
frequencies. 

Table I gives the atmospheric pressure value assumed for the
dielectric constant\cite{polian,kuroda}, the pressure 
coefficient of the phonons\cite{clemenz}
and the reference from which they were taken. 
Once the behaviour under pressure of all the parameters is known,
we can calculate the mobility for 3D electrons ($\mu_3$) by
considering different compensation ratio ($x$). Figure 8 
shows the results of this calculation.

As regards the 2D electron, according to the model proposed 
in Reference \cite{18}, the electron mobility determined 
by homopolar optical phonon scattering is much 
smaller than the 3D electron mobility due to the higher value of
the effective electron-phonon coupling constant ($g_{ef}^2$), that 
depends on the localization parameter of 2D electrons along the 
c-axis ($b_0$):

\begin{eqnarray}
{g_{ef}^2}=\frac{3\pi}{8}\sqrt{\frac{\hbar^2 {b_0}^2}
{2 {m^*_{\perp}}^2 \hbar \omega_{hp}}} g^2\quad,
\end{eqnarray}

where

\begin{eqnarray}
b_{0}=\sqrt{\frac{33 e^2 m^*_{\parallel}n_s}
{8\epsilon_{0\parallel}\hbar^2}}
\end{eqnarray}

\noindent
and where $n_s$ is the density of 2D electron per unit area in the
planar defects. As we pointed out above,
the pressure dependence of the involved quantities are well known 
and can be taken from literature.  The mobility for
the 2D electrons ($\mu_2$) can be calculated in the relaxation
time approximation and the results are also shown in Figure 8. 
The higher relative increase of  2D electron  mobility with
respect to that of 3D electrons is mainly due to the decrase of 
the localization parameter $b_0$, as $\epsilon_{0\parallel}$ 
increases under pressure. 

It is to be outlined that the model proposed in Reference \cite{18} 
to explain the temperature dependence of electron mobility in a 
large variety of InSe samples can also explain the pressure effects. 
In the framework of this model, once computed separately the pressure
dependence of the 2D mobility ($\mu_2$) and 3D mobility ($\mu_3$), we 
have calculated the pressure dependence of the Hall mobility and 
the apparent electron concentration by means of\cite{mir}:

\begin{eqnarray}
\mu=\frac{n_2 \mu_2^2 + n_3 \mu_3^2}
{n_2 \mu_2 + n_3 \mu_3}
\end{eqnarray}

\hspace{1in}and

\begin{eqnarray}
\quad n=\frac{(n_2 \mu_2+n_3 \mu_3)^2}
{n_2 \mu_2^2+n_3 \mu_3^2} \quad.
\end{eqnarray}

Taking $n_2$, $n_3$ and $x$ at room pressure as fitting parameters, 
we fit Equations (10) and (11) to the experimental results. Fits
are shown in Figure 9 and the agreement with the experimental
results is quite good. In the Sn-doped samples the
decrement of $n_3$ under pressure, due to the presence of the
trapping centers, makes that $n_3$ becomes
smaller than $n_2$ at around 20 kbar. As 2D electrons have a 
smaller mobility than
the 3D ones (see Fig. 8), the decrease of $n_3$ is then 
responsible for the drop of the Hall mobility. Instead of that, 
in Si-doped samples such a strong decrease of the mobility is not 
observed because $n_2$ is of the same order that
$n_3$. In this way, the behaviour of the effective electron 
concentration and the mobility are less sensitive to the decrease of 
$n_3$. In fact, the slight increase of the carrier concentration 
observed in some samples can be also accounted for. If $n_2 \ge n_3$, 
at ambient pressure, $n$ is closer to $n_3$ due to the higher mobility 
of the 3D electrons. As $n_3$ decreases under compression, $n$ tends 
to $n_2$ and then slighty increases.

\subsection{Phase transition and precursor effects}
\indent

Resistivity measurements at high pressure can give information 
about the transition from the layered structure to the NaCl-type 
metallic phase, that was detected at about 100 kbar through x-ray
diffraction and reflectivity measurements\cite{3}. In Figure 3 a 
resistivity drop is observed that can be related to this 
semiconductor-metal 
transition.  Nevertheless, resistivity measurements show that 
some irreversible changes occur in the material at much lower  
pressures. The resistivity decrease starting at about 45 kbar 
corresponds to the thermopower drop 
that is also observed above that pressure.  
Both process are nonreversible and suggest that there is 
a pressure above which some kind of structural instability 
occurs, leading to the creation of a large concentration of donor 
centers that make the material to 
show a degenerate behaviour. Other authors have observed 
that above 60 kbar the sample
becomes irreversibly opaque\cite{clemenz,9}. From reflectivity 
measurements\cite{3}, it has been also shown that 
a plasma reflection structure is observed at 80 kbar, i.e., 20 
kbar below the transition to the metallic phase. If we assume 
that the reflectivity minimum, that occurs at  $\omega_s$=0.8 eV,  
corresponds to the plasma frequency, then the electron
concentration (for an effective mass of the order of $m_o$) would 
be of the order of 10$^{21}$ cm$^{-3}$. Results of Figure 4 
($\circ$) show that the resistivity decrease is due to the sharp
increase of the electron concentration. The highest measured 
electron concentration is of the order of 10$^{18}$ cm$^{-3}$. As 
the resistivity still lowers by three orders of magnitude, the 
electron concentration would be of the same order of that 
estimated from the value of the plasma 
frequency, which is also coherent with the low value of the 
thermopower.

In the metallic phase, if one assumes an electron per unit cell, the 
electron concentration would be of the order of 2.5 $\times$ 
10$^{22}$ cm$^{-3}$. It means that before the transition the defect 
concentration is very high, 
of the order of 1 defect per  20 to 25 unit cells. 
In this range of pressure the layered phase seems to be stable
only when it contains a high defect concentration. It is reasonable 
to conceive those defects as precursors of the phase transition.  
The atom plane sequence of the layer phase (Se-In-In-Se) must 
change to the Se-In-Se-In sequence of the NaCl-type phase. A 
precursor defect of the high 
pressure plane sequence  would be one in which a In
atom shifts from the intralayer position to an interlayer 
octahedral site, typical of the NaCl-type structure. Raman 
measurements under pressure show that in layered crystals from 
the III-VI family  there is a charge 
tranfer from the In-In intralayer bonds to the interlayer space, 
which increases the overall ionicity of the crystal
under pressure. These results indicate that the In-In intralayer
bond tends to be relatively weakened\cite{2,5}.
A possible precursor defect that fills all the conditions would be 
that represented in Figure 10, in which an In-In bond is broken and 
one of the In atoms jumps to the octahedral interlayer site, giving 
rise to 2 dangling cation bonds.
In partly ionic compounds, cation dangling bond occur in anion
vacancies that have donor character. It seems then to be
reasonable to atribute donor character to the defect here
proposed. 

\section{Conclusions}
\indent

Systematic Hall effect, resistivity and thermopower measurements 
have been carried out in Sn-doped and Si-doped and NTD InSe. The 
strong decrement of the electron concentration under compression in 
samples exhibiting a 3D behaviour is explained through the 
existence of an excited minimum in the conduction band, which shifts 
towards lower energies with a rate of -105 meV/GPa as pressure 
increases, and whose related impurity level traps electrons as it 
approches the Fermi level. The behaviour of transport parameters 
under pressure up to 40 kbar in different samples has been explained 
through a model in which both 3D and 2D electrons has been taken 
into account. The fit of this model to the experimental results 
is quite satisfactory. In addition, around 50 kbar a sharp increase 
of the electron concentration was detected. Above this pressure both 
resistivity and thermopower changes start to be nonreversible, this 
is an evidence of a pressure-induced structural instability of InSe. 
These phenomena have been related with the presence of defects which 
are precursor of the phase transition to the NaCl metallic phase. 
This transition has been observed as a drop in the resistivity 
at around 105 kbar. 

\vspace{0.3in}
\noindent
{\Large{\bf Acknowledgments}}

\vspace{0.1in}

This work was supported by the Spanish Government CICYT under 
Grant No MAT95 - 0391 and by Generalitat Valenciana under
Grants No GV - 2205/94 and No GV - 3235/95.

\pagebreak
\noindent
{\Large {\bf Table captions}}

\vspace{0.3in}
\noindent
{\bf Table I}: Dielectric constants at room pressure and pressure
dependence of the phonon frequencies.

\pagebreak
\noindent
{\Large {\bf Figure captions}}

\vspace{0.3in}
\noindent
{\bf Figure 1}: Resistivity as a function of hydrostatic pressure 
for several n-type InSe samples. The lines are just guides to
the eye.

\vspace{0.15in}
\noindent
{\bf Figure 2}: (a) Electron concentration and (b) mobility as a 
function of hydrostatic pressure for Sn ($\Box$ and $\triangle$) 
and Si ($\circ$ and $\Diamond$) doped and NTD ($\bullet$) InSe 
samples.

\vspace{0.15in}
\noindent
{\bf Figure 3}: Resistivity as a function of pressure for a 
Sn-doped InSe sample in quasi-hydrostatic conditions. Curves 1 were 
taken during upstroke and curves 2 during downstroke. The inset shows
the experimental assembly. 

\vspace{0.15in}
\noindent
{\bf Figure 4}: Resistivity, electron concentration and 
mobility as a function of pressure for several 
Sn-doped InSe samples under quasi-hydrostatic conditions.

\vspace{0.15in}
\noindent
{\bf Figure 5}: Same as Fig. 4 for two Si-doped InSe 
samples under quasi-hydrostatic conditions.

\vspace{0.15in}
\noindent
{\bf Figure 6}: Thermopower as a function of pressure for one Sn doped
sample in quasi-hydrostatic conditions. The inset shoow the
experimental assembly.

\vspace{0.15in}
\noindent
{\bf Figure 7}: Change under pressure of the Fermi level ($E_F$)
and the deep impurity level ($E_T$) with regards to the absolute
minimum of the conduction band ($E_C$) ($\circ$, $\Box$) and to the
bottom of the 2D subbands ($E_{C2}$) ($\bullet$).  

\vspace{0.15in}
\noindent
{\bf Figure 8}: calculated $\mu_2$ (6) and $\mu_3$ (1 to 5) by 
assuming different concentration ratio  and
taking $n_3$ = 8 10$^{16}$ cm$^{-3}$ and $n_2$ = 5 10$^{15}$ 
cm$^{-3}$. (1) no compensation, (2) 20 $\%$, (3) 40 $\%$, (4) 60
$\%$ and (5) 80 $\%$. The dashed line gives the result of a fit
to the data points by using a second order polinomial (see Eq.
(3)). The solid lines are just guides to the eye.

\vspace{0.15in}
\noindent
{\bf Figure 9}: Fitting through the model proposed in section 4
to the experimental (a) Hall mobility and (b) electron concentration.

\vspace{0.15in}
\noindent
{\bf Figure 10}: Schematic view of the possible precursor defect
originated by the jump of an In atom from the intralayer space (a)
to the interlayer space (b). ($\circ$) In atoms
and ($\bullet$) Se atoms. 

\pagebreak

{\large
\begin{table}
\begin{itemize}
\item[]\begin{tabular}{@{}lcccccc}
\hline
&$\epsilon_{\infty\perp}$&$\epsilon_{\infty\parallel}$&
$\epsilon_{0\perp}$&
$\epsilon_{0\parallel}$&$\delta \omega_{hp}/\delta P$&
$\delta \omega_{LO}/\delta P$\\
&&&&&(cm$^{-1}$/kbar)&(cm$^{-1}$/kbar)\\
\hline
&7.4&7&10.9&7.8&0.54 &
0.36\\
Reference&40&40&41&41&7&7\\
\hline
\end{tabular}
\end{itemize}
\end{table}}

\vspace{0.5in}
\begin{thebibliography}{50}

\bibitem{1}A. Polian, J.M. Besson, M. Grimsditch and H. Vogt, 
Phys.Rev.B {\bf 25}, 2767 (1982).

\bibitem{2}M. Gauthier, A. Polian, J.M. Besson and A. Chevy, 
Phys.Rev.B {\bf 40}, 3837 (1989).

\bibitem{3}U. Schwarz, A.R. Go\~ni, K. Syassen, A. Cantarero and 
A. Chevy, High Pressure Research {\bf 8}, 396 (1991).

\bibitem{4}A. Polian, J.C. Chervin and J.M. Besson,
Phys.Rev.B {\bf 22}, 3049 (1980).

\bibitem{5}N. Kuroda, O. Ueno and Y. Nishina,
Phys.Rev.B {\bf 35}, 3860 (1987).

\bibitem{6}J.M. Besson, K.P. Jain and A. Khun,
Phys.Rev.Lett. {\bf 32}, 936 (1974).

\bibitem{clemenz}C. Ulrich, M.A. Mroginski, A.R. Go\~ni, A. Cantarero,
U. Schwarz, V. Mu\~noz and K. Syassen,{\it Proceedings of the
Seventh International Conference on High Pressure Semiconductor 
Physics} [PSS (a) {\bf 198}, 121 (1996)].

\bibitem{7}M. Mejatty, A. Segura, R. Le Toullec, J.M. Besson, 
A. Chevy and H. Fair, J. Phys. Chem. Solids {\bf 39}, 25 (1978).

\bibitem{8}N. Kuroda, O. Ueno and Y. Nishina,
J.Phys.Soc.Jpn. {\bf 55}, 581 (1986).

\bibitem{9}A.R. Go\~ni, A. Cantarero, U. Schwarz, K. Syassen and  
A. Chevy, Phys.Rev.B {\bf 45}, 4221 (1992).

\bibitem{10}N. Kuroda, Y. Nishina, H. Ywasaki and Y. Watanabe, 
Solid State Commun. {\bf 38}, 139 (1981).

\bibitem{11}A.A. Ismailov, Sh.G. Gasymov, T.S. Mamedov and K.R. 
Aliakhverdlev, Fiz. Tekh. Poluprovdn. {\bf 26}, 1995 (1992)
[Sov. Phys. Semicond. 26, 1122 (1993)].

\bibitem{18}A. Segura, B. Mar\'{\i}, J.P. Mart\'{\i}nez-Pastor 
and A. Chevy, Phys.Rev.B {\bf 43}, 4953 (1991).

\bibitem{12}A. Chevy, J. Cryst. Growth {\bf 67}, 119 (1984).

\bibitem{13}A. Chevy, J.Appl.Phys. {\bf 56}, 978 (1984).

\bibitem{14}B. Mar\'{\i}, A. Segura and A. Chevy, 
Appl.Surf.Sci. {\bf 50}, 415  (1991).

\bibitem{15}R. Pareja, R. de la Cruz, B. Mar\'{\i}, A. Segura, 
V. Mu\~noz, Phys.Rev.B {\bf 47}, 2870 (1993).

\bibitem{16}J.L. Van der Pauw, Philips Research Report {\bf 13},
1 (1958).

\bibitem{17}J.M. Besson, G. Hamel, P. Grima, R.J. Nelmes,
J. Loveday, S. Hull and D. H\"ausermann, High Pressure Research 
{\bf 8}, 625 (1992).

\bibitem{piro}M. Nishikawa and S. Akimoto, High Temperature - High
Pressure {\bf 3}, 161 (1971).

\bibitem{19}J. Riera, A. Segura and A. Chevy,
Appl.Phys.B {\bf 54}, 428 (1992).

\bibitem{20}J. Mart\'{\i}nez-Pastor, A. Segura, C. Julien, A. Chevy, 
Phys.Rev.B {\bf 46}, 4607 (1992).

\bibitem{21}A. Segura and A. Chevy, Phys.Rev.B {\bf 49}, 4601 (1994).

\bibitem{yo}D. Errandonea [to be published].

\bibitem{band1}A.Bourdon, A.Chevy and J.M.Besson, {\it
Proceedings of the Fourteenth International Conference of
Physics of Semiconductors} [Int. Phys. Conf. Ser. {\bf 43}, 1371
(1979)].

\bibitem{band2}C. Ulrich, A.R. Go\~ni, K. Syassen, O. Jepsen,
A. Cantarero and V. Mu\~noz, {\it Proceedings of the XV AIRPAT} 
\& {\it XXXII EHPRG Conference},  edited by 
W. Trzeciakowski (World Scientific, 1996), p. 411.

\bibitem{kress1}J.C. Portal, R. Nicholas, E. Kress-Rogers, A. Chevy,
J.M. Besson, J. Galibert and D. Perrier, {\it Proceedings of the
Fifteenth International Conference of Physics of Semiconductors}
[J. Phys. Soc. Jpn. Suppl. A {\bf 49}, 879 (1980)].

\bibitem{portal}E. Kress-Rogers, R. Nicholas, J.C. Portal and
A. Chevy, Solid State Commun. {\bf 44} 379 (1982).

\bibitem{kress2}R. Nicholas, E. Kress-Rogers, J.C. Portal,
J. Galibert and A. Chevy, Surf. Sci. {\bf 113}, 339 (1982).

\bibitem{kress3}E. Kress-Rogers, G. Hoppert, R. Nicholas, W. Hayes,
J.C. Portal and A. Chevy, J.Phys. C {\bf 16}, 4285 (1983).

\bibitem{kress4}E. Kress-Rogers, R. Nicholas and A. Chevy, J.Phys.
C {\bf 16}, 2439 (1983).

\bibitem{howell}D.F. Howell, R. Nicholas, C. Langerak,
J. Singleton, T. Janssen and A. Chevy, J.Phys. Condens.Matter {\bf
1}, 7493 (1989).

\bibitem{stern}F. Stern and W.E. Howard, Phys.Rev. {\bf 163}, 
816 (1967).

\bibitem{nag}B.R. Nag, {\it Electron Transport in Compound
Semiconductors}, Springer Series in Solid State Sciences {\bf
vol.11} (Springer - Verlag, Berlin Heidelberg 1980).

\bibitem{fivaz}R. Fivaz and E. Moser, Phys.Rev {\bf 163}, 743 (1967).

\bibitem{brooks}H. Brooks, {\it Advances in Electronics and
Electron Physics} {\bf vol.7} (Academic, New York, 1955).

\bibitem{gasanly}N.  Gasanly, B. Yavadov, V. Tagirov and E.
Vinogradov, Phys.Stat.Sol. (b) {\bf 89}, k43 (1978).

\bibitem{lang}I.G. Lang and U. Pashabekova, Fiz.Tverd.Tela {\bf
6}, 3640 (1965) [Sov.Phys.-Solid State {\bf 6},2913 (1965)].

\bibitem{kane}E.M. Conwell, {\it High Field Transport in
Semiconductors}, (Academic, New York, 1972).

\bibitem{polian}N. Piccioli, R. Le Toullec, F. Bertrand and
J.C. Chervin, J.Phys. (Paris) {\bf 42}, 1129 (1981).

\bibitem{kuroda}N. Kuroda and Y. Nishima Solid State Commun. {\bf
34}, 481 (1980).

\bibitem{mir}P. Kir\`eev, {\it La Physique des Semiconducteurs}
(Mir, Moscow, 1975).

\end{thebibliography}
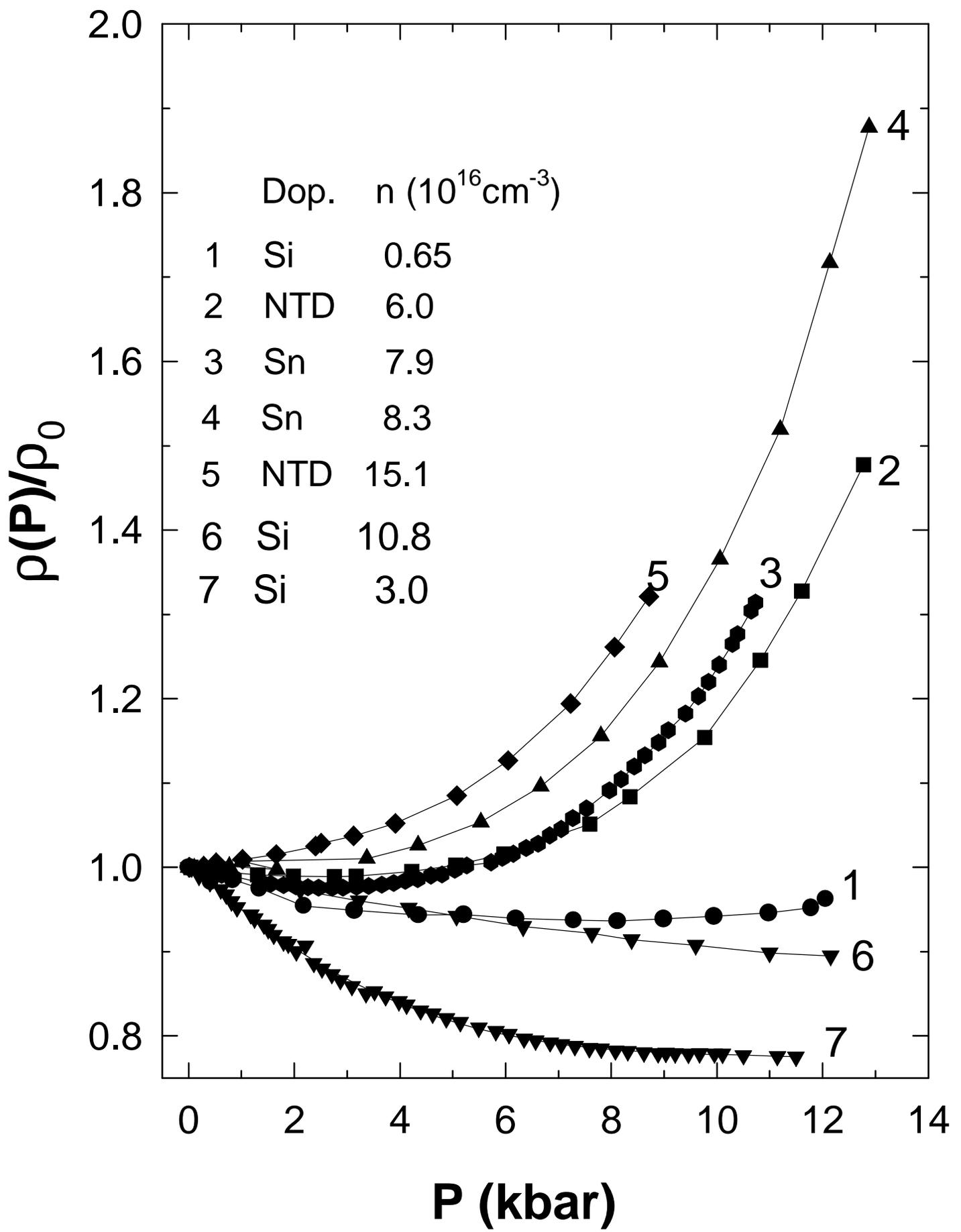
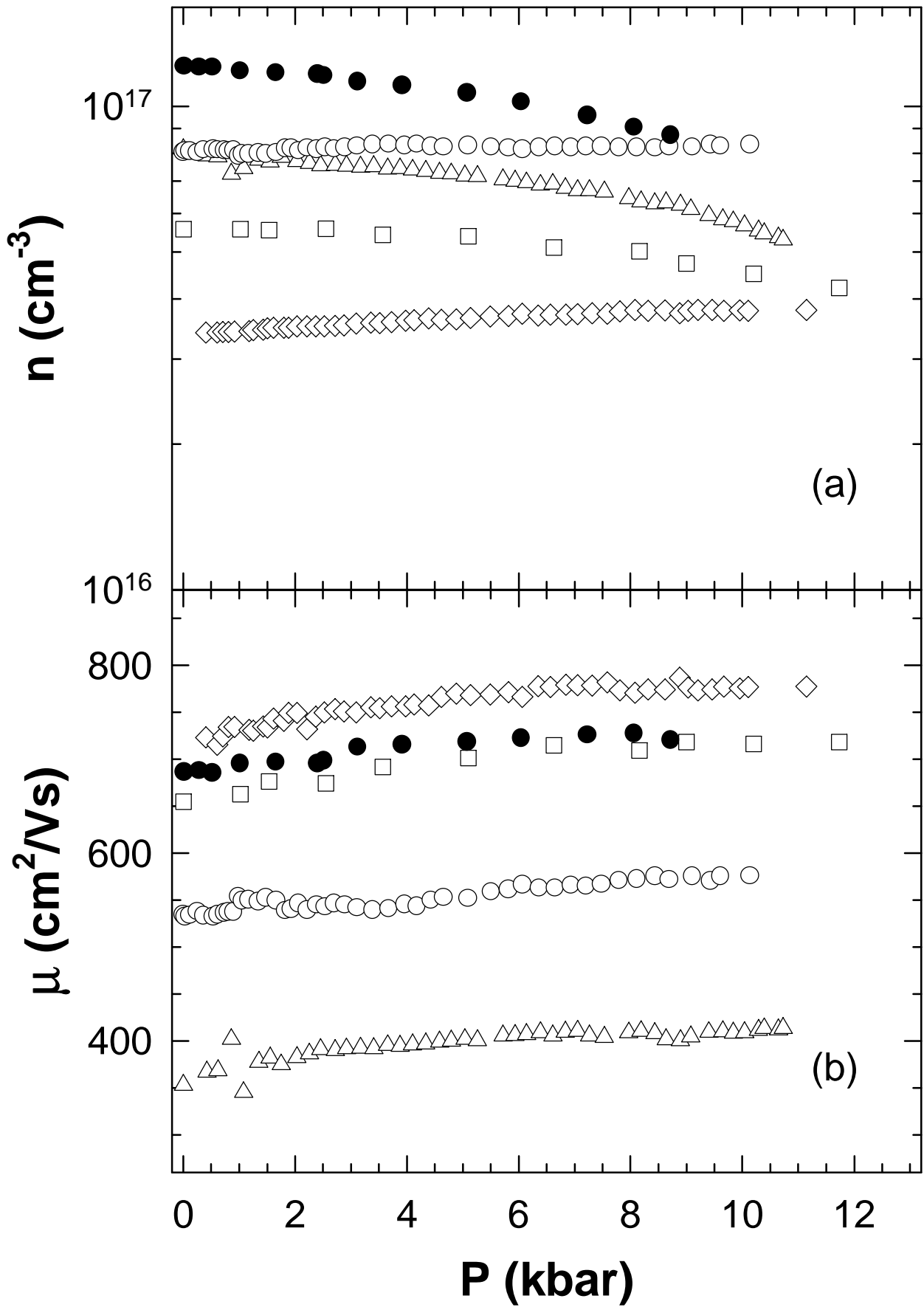
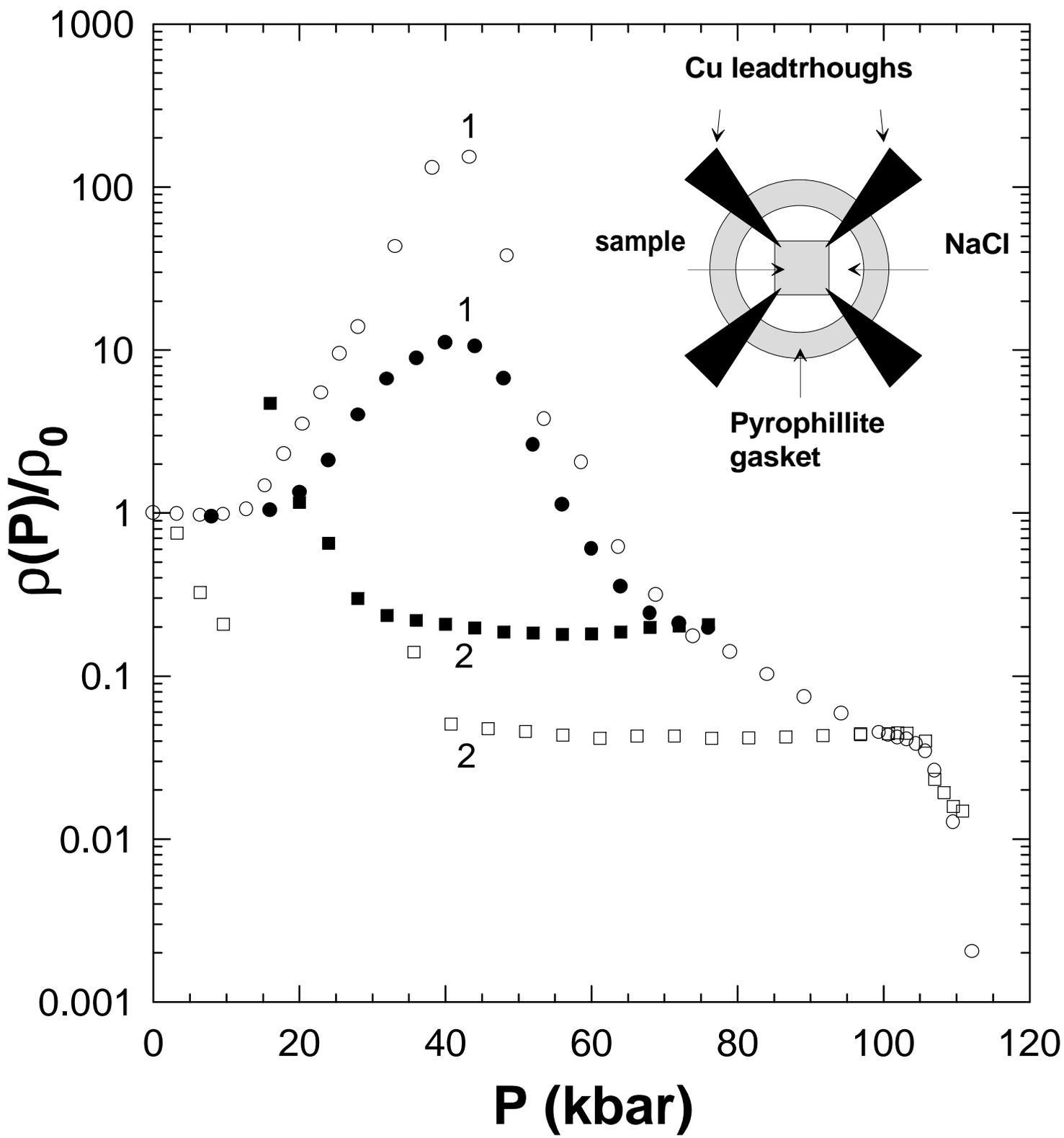
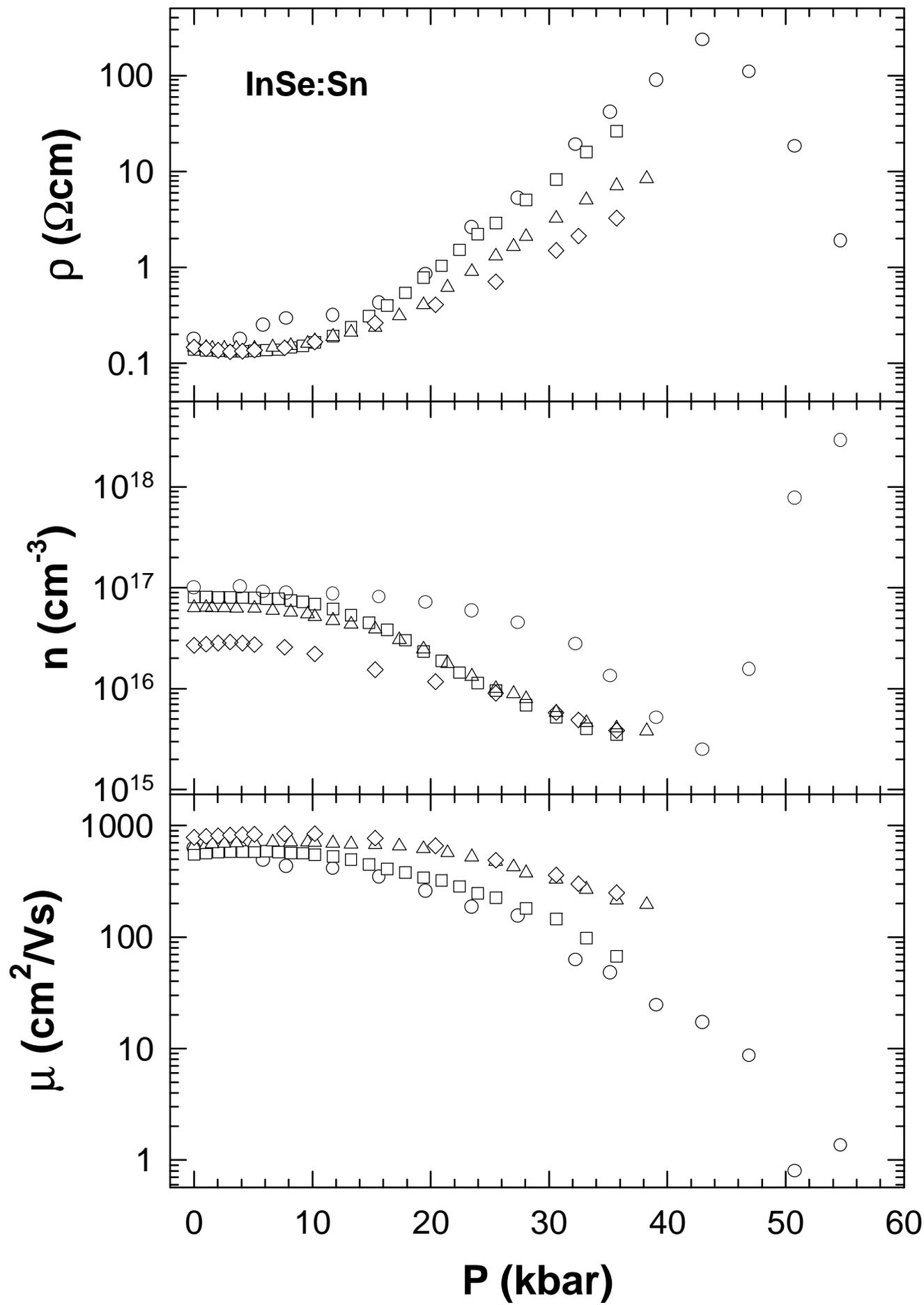
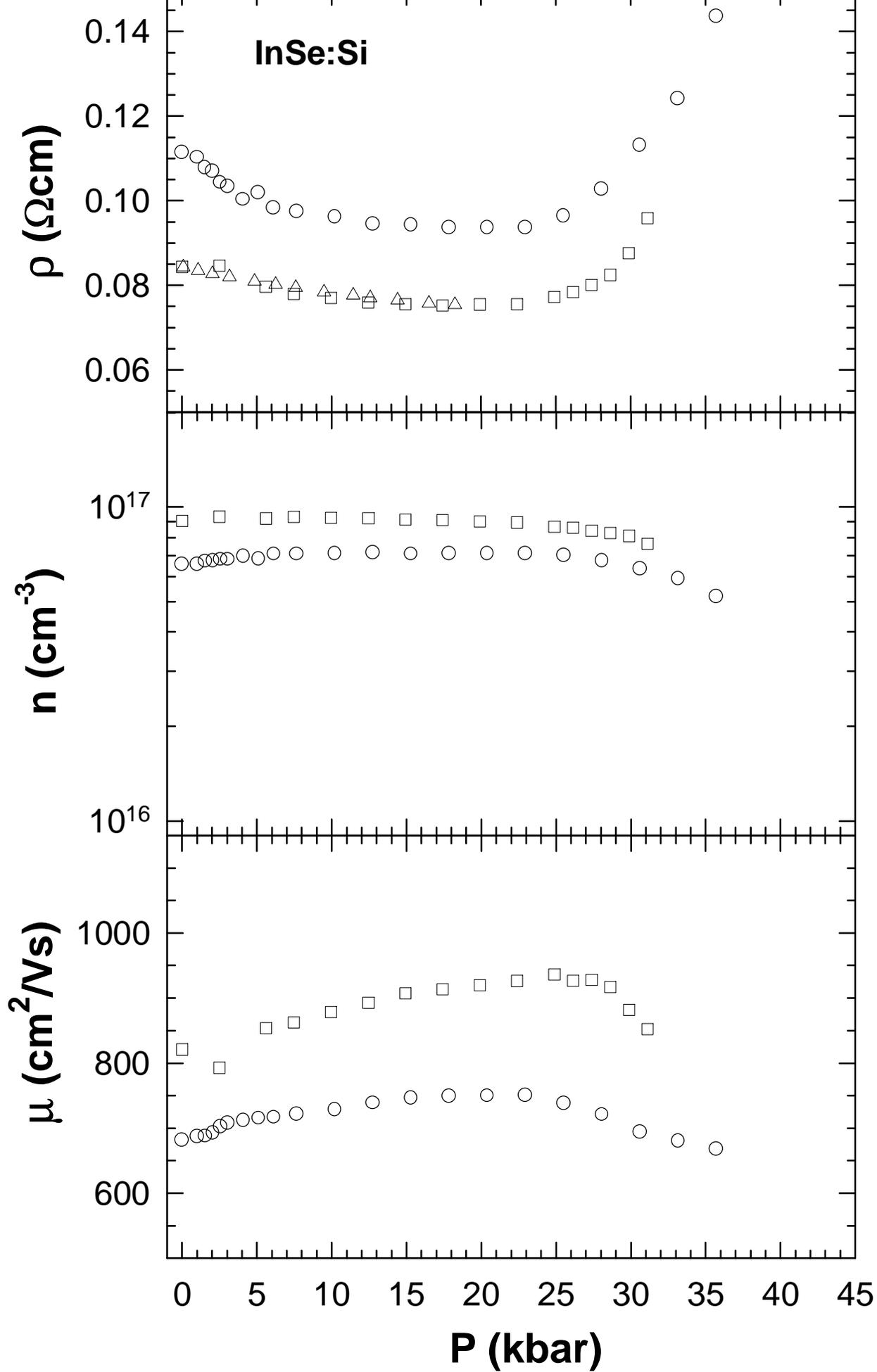
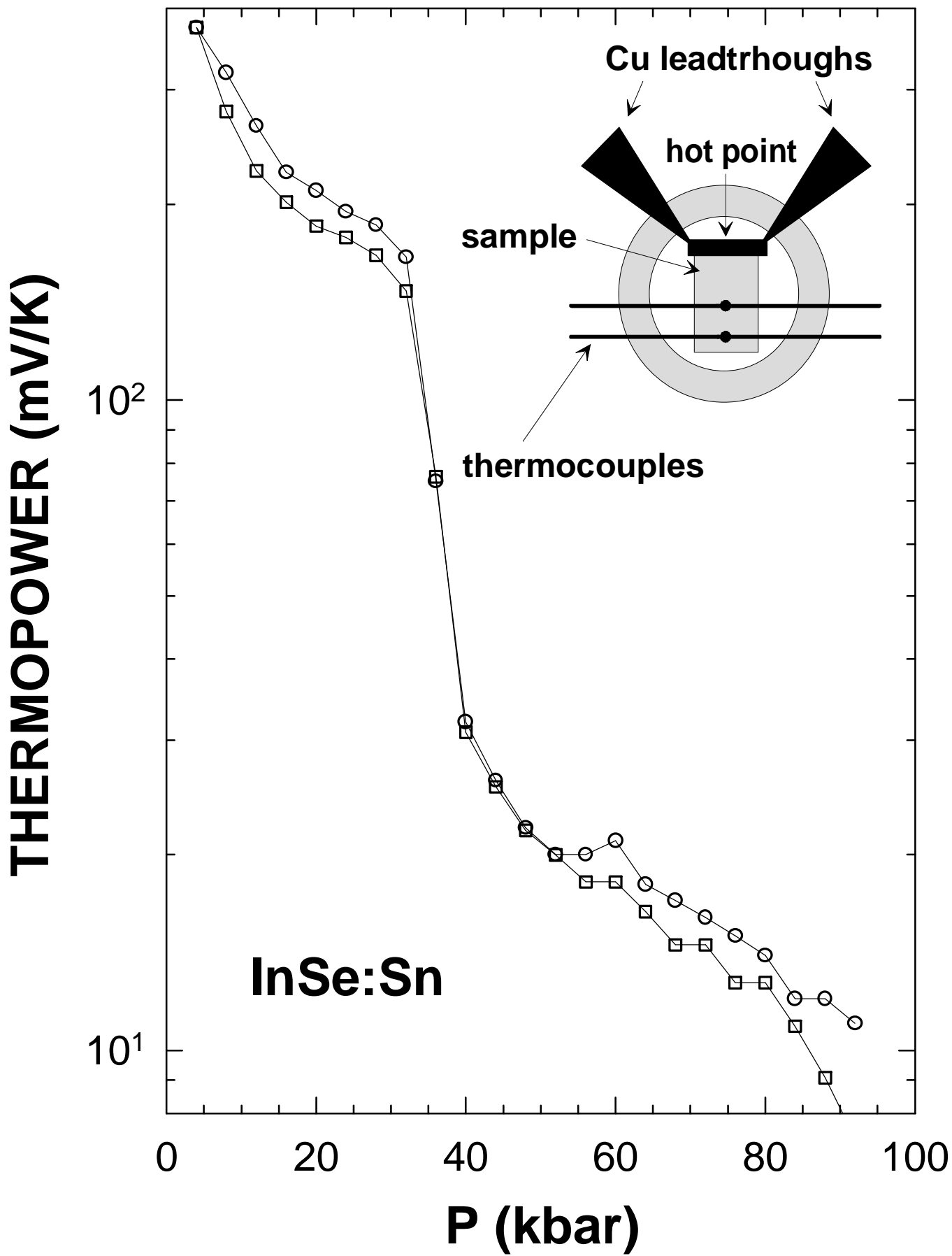
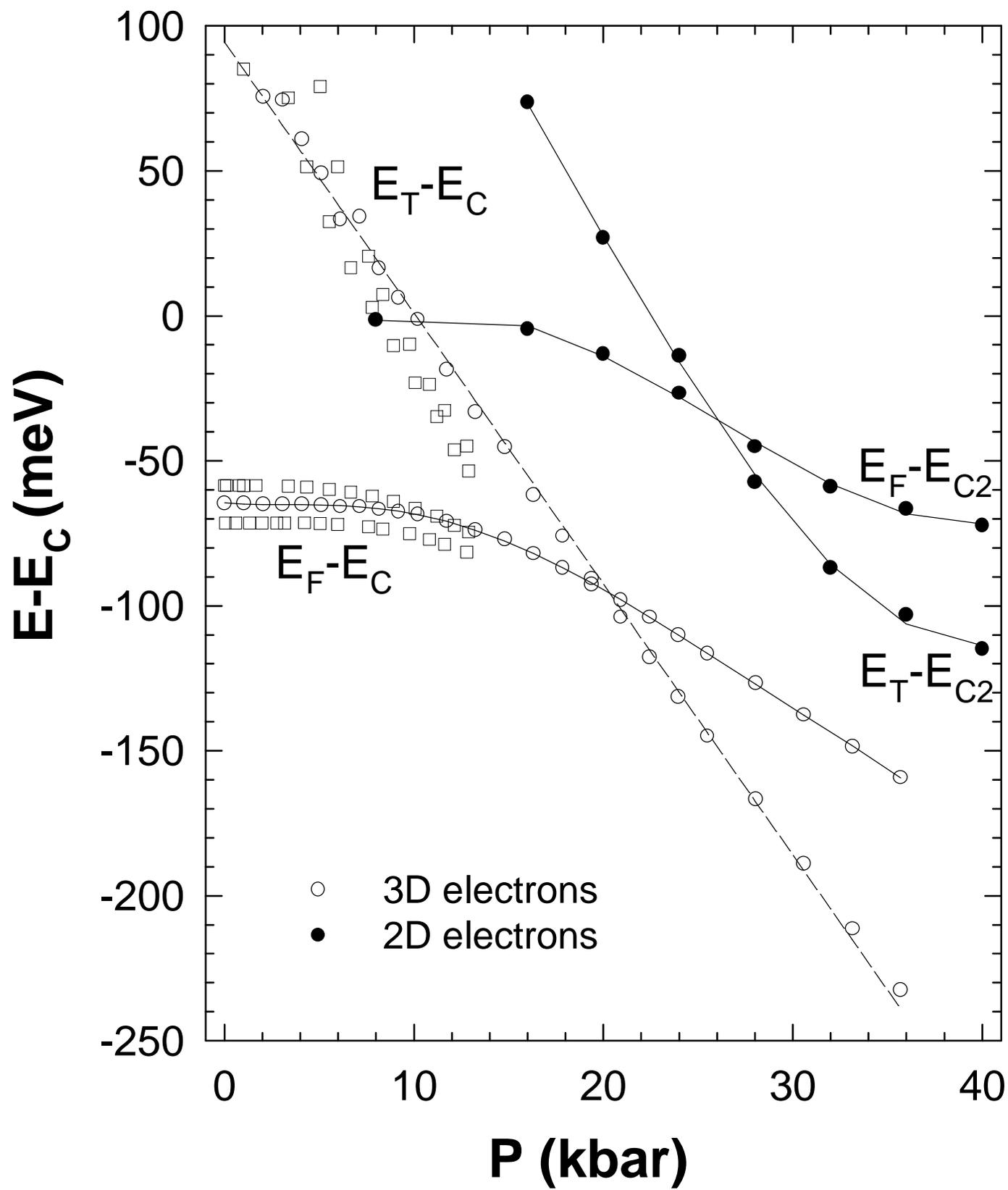
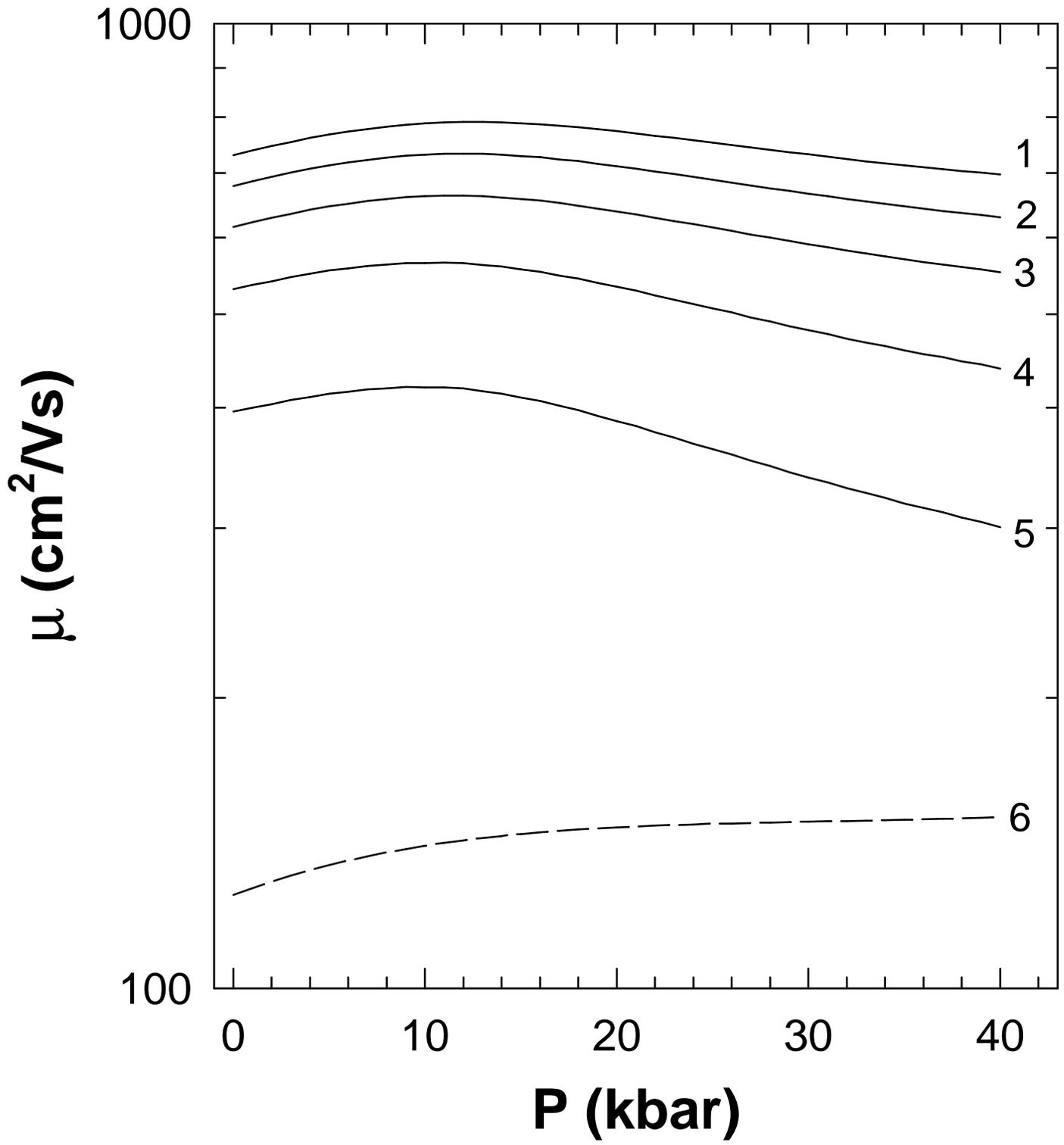
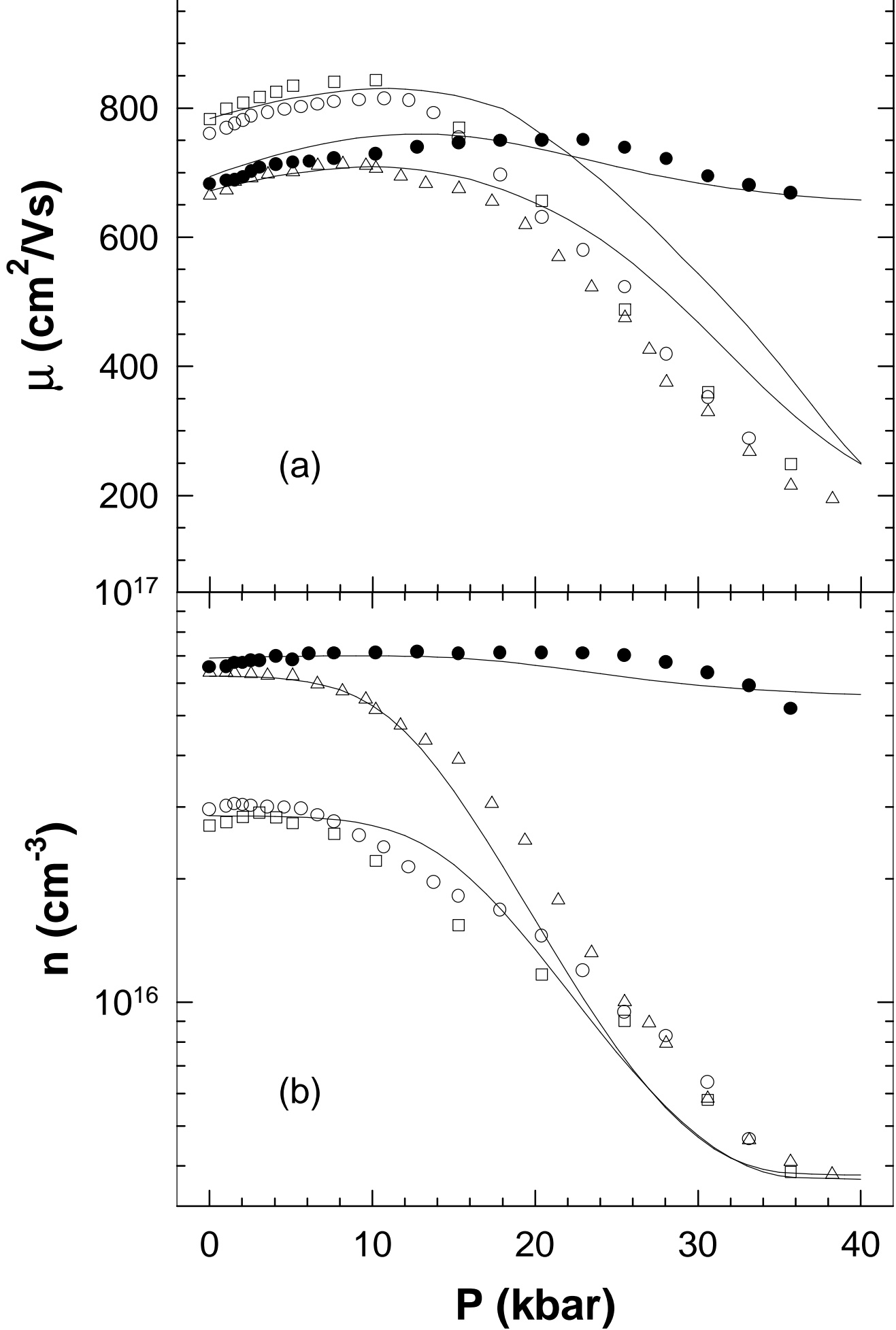
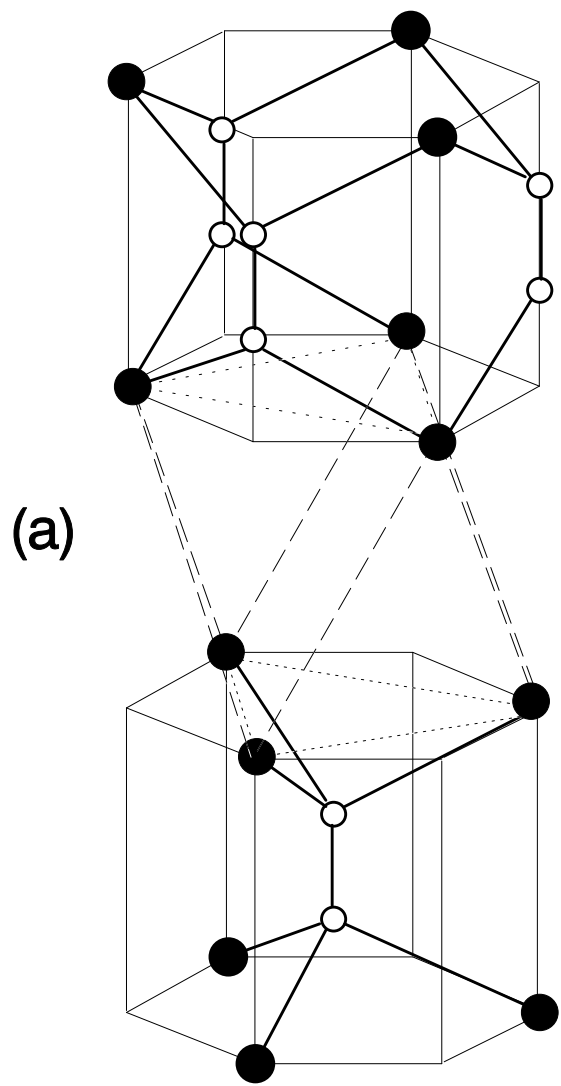
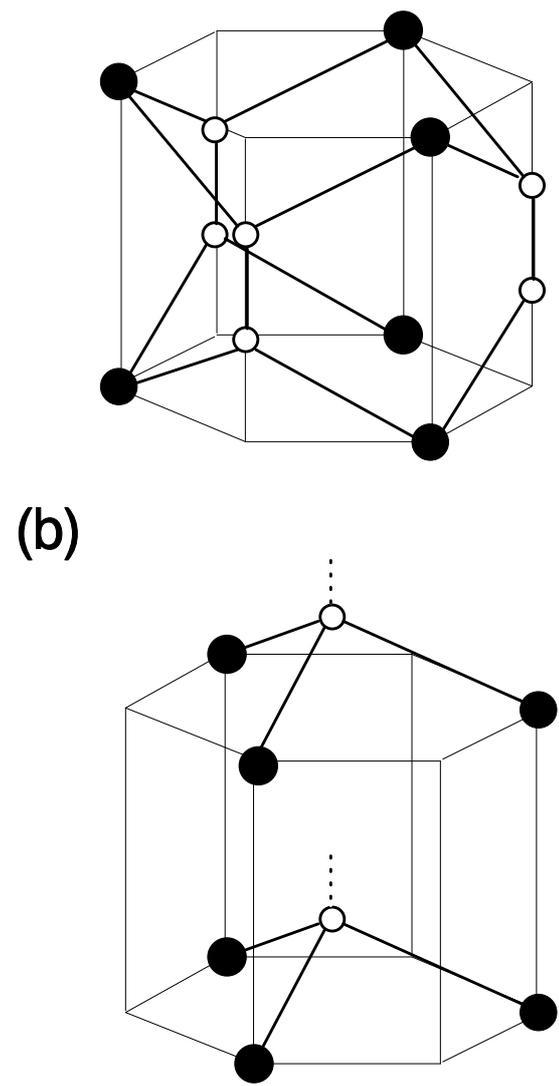
\end{document}